\documentclass[a4paper]{article}
\usepackage{graphicx,supertabular}
\title{New list of peculiar velocities of RFGC galaxies}
\author{S. L. Parnovsky and A. V. Tugay}
\newcommand{\E}{\rlap{E}}
\newcommand{\B}{\rlap{B}}
\newcommand{\N}{\rlap{N}}

\begin{document}
\maketitle
\noindent
Astronomical Observatory of Kyiv Taras Shevchenko National
University, Observatorna 3, 04053, Kyiv, Ukraine

\begin{abstract}
We present a new version of the list of peculiar velocities of
1561 flat edge-on spirals from the RFGC catalogue. It differs from
the previous version by 233 new data and 34 corrected data.
A new regression was used for distances
estimation based on the Tully-Fisher relationship in the ``linear
diameter -- HI line width'' variant. Moreover, we present
velocities for 3 models of galaxies collective motion. They are a
D-model (dipole, Hubble expansion + bulk motion with constant
speed), DQ-model (a quadrupole terms are added) and DQO-model
(DQ-model + octopole).
\\

\end{abstract}

\section{Brief history}

The study of non-Hubble motion is very important for cosmology and
cosmography. So we need large representative samples of galaxies
peculiar velocities, covering the whole sky. Karachentsev (1989)
proposed to use thin edge-on spiral galaxies as ``test
particles'' for collective non-Hubble motion of galaxies. He was
the head of the group of astronomers from the Special
Astrophysical Observatory of the Russian Academy of Sciences
(Russia) and Astronomical Observatory of the Kyiv Taras Shevchenko
National University (Ukraine) who prepared the catalogues of such
galaxies -- Flat Galaxies Catalogue (FGC) (Karachentsev et al, 1993)
and its revised version (RFGC) (Karachentsev et al, 1999).
Preparation included special all-sky search
for late edge-on spiral galaxies and selection of objects
satisfying the conditions $a/b \ge7$ and $a \ge0.6'$, where $a$
and $b$ are the major and minor axes. The FGC and RFGC catalogues
contain 4455 and 4236 galaxies respectively and each of them
covers the entire sky. Since the selection was performed using the
surveys POSS-I and ESO/SERC, which have different photometric
depth, the diameters of the southern-sky galaxies were reduced to
the system POSS-I, which turned out to be close to the system
$a_{25}$. The substantiation of selecting exactly flat galaxies
and a detailed analysis of optical properties of the catalogue
objects are available in the texts of FGC, RFGC and in references therein.

By 2001 we had information about radial velocities and HI 21\,cm
line widths, $W_{50}$, or rotational curves $V_{rot}$ for 1327
RFGC galaxies from different sources listed below. Some of them
are obvious outliers. After omitting these ``bad'' data the sample
was reduced to 1271 galaxies (see Fig. 1). These galaxies lie
quite homogeneously over the celestial sphere except a Milky Way
zone (see Fig. 2). This sample was the basis for building a
regression for estimation of galaxies' distances. In the paper
(Parnovsky et al, 2001) three regressions were obtained for different
models of peculiar velocity field. The regression for simplest D-model
was used to create a list of peculiar velocities of 1327 RFGC galaxies
(Karachentsev et al, 2000). At
the same time, various measurements of radial velocities and HI
21\,cm line widths were carried out. Few years later the HyperLeda
extragalactic database ( http://leda.univ-lyon1.fr) contained some
new data for RFGC galaxies. Taking into account these data we
compiled a new sample of 1561 RFGC galaxies with known radial
velocities and HI 21\,cm line widths [6]. It contains 233 new
data. Data for another 34 galaxies were changed because their
HyperLeda data fitted regression much better than previous data.
After discarding 69 ``bad'' data we got a sample of 1492 data,
which was used for obtaining regressions for estimation of
distances (Parnovsky and Tugay, 2004). We used the same three models
of collective motion as in the paper (Parnovsky et al, 2001).

\begin{figure}[t]
\centering
\includegraphics[width=12cm]{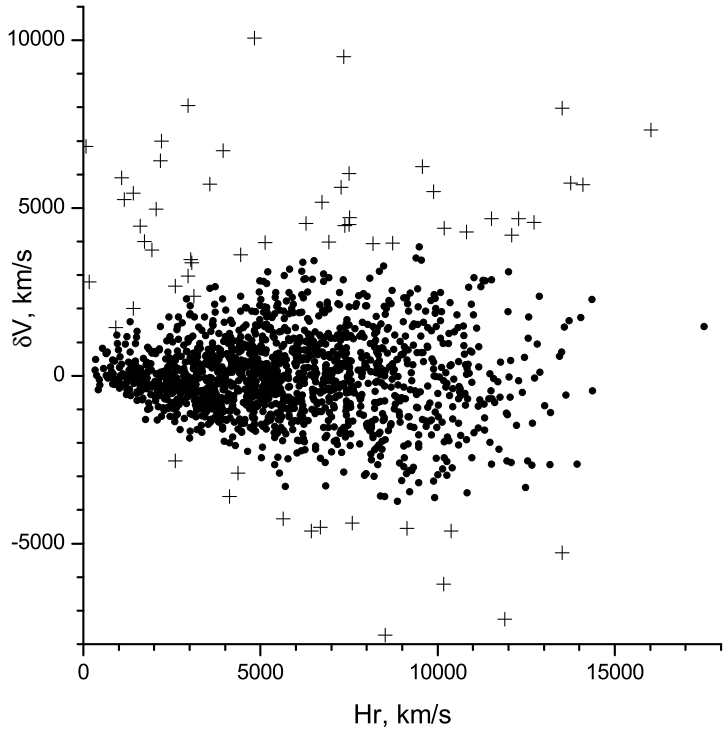}
\caption{Deviations of radial velocities from regression for D-model vs distances. Crosses
mark ``bad'' data.}
\end{figure}
\begin{figure}[t]
\centering
\includegraphics[width=12cm]{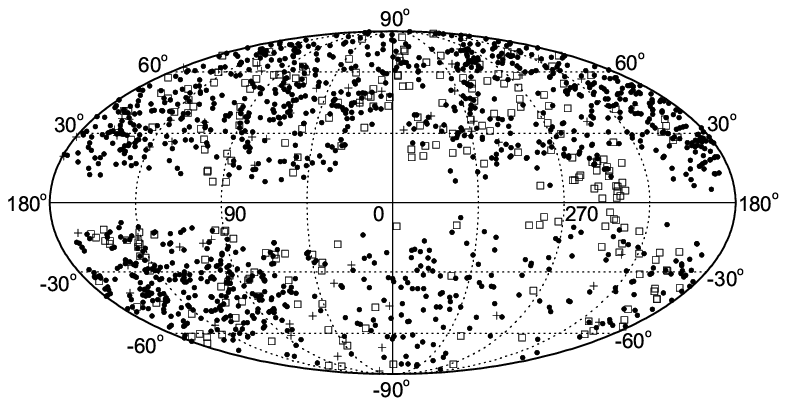}
\caption{Distribution of 1561 flat galaxies over the celestial sphere in the galactic coordinates. Crosses
mark ``bad'' data, squares -- new entries.}
\end{figure}

\section{Models of collective motion of galaxies}

In the paper (Parnovsky et al, 2001) the velocity field was expanded in terms
of galaxy's radial vector $\vec{r}$. It was used to obtain some
models of dependence of galaxy's radial velocity $V$ from
$\vec{r}$. In the simplest D-model (Hubble law + dipole) we have

\begin{equation}
V=R+V^{dip}+\delta V,R=Hr, V^{dip}=D_{i}n_{i},
\vec{n}=\vec{r}/r,r=|\vec{r}|,
\end{equation}

\noindent where $H$ is the Hubble constant, $\vec{D}$ is a
velocity of homogeneous bulk motion, $\delta V$ is a random
deviation and $\vec{n}$ is a unit vector towards galaxy. In our
notation we use the Einstein rule: summation by all the repeating
indices. After addition of quadrupole terms we obtain a DQ-model

\begin{equation}
V=R+V^{dip} + V^{qua}+\delta V, V^{qua}=RQ_{ik} n_{i} n_{k}
\end{equation}

\noindent with symmetrical traceless tensor $\bf{Q}$ describing
quadrupole components of velocity field. The DQO-model includes
octopole components of velocity field described by vector
$\vec{P}$ and traceless symmetrical tensor $\bf{O}$ of rank 3:

\begin{equation}
V=R+V^{dip} + V^{qua} + V^{oct}+\delta V, V^{oct}
=R^{2}(P_{i}n_{i} +O_{ikl} n_{i} n_{k} n_{l}).
\end{equation}

In order to calculate a peculiar velocity $V^{pec}=V-Hr$ one must
have an estimation of galaxy's distance $r$ or a corresponding
Hubble radial velocity $R=Hr$. We use a generalized Tully-Fisher
relationship (Tully and Fisher, 1977)in the ``linear diameter -- HI line width'' variant.
It has a form (Parnovsky et al, 2001, Parnovsky and Tugay, 2004)

\begin{eqnarray}
R & = &(C_1+C_2B+C_3BT)W_{50}/a_r+C_4W_{50}/a_b \nonumber \\ & &
{}+C_5(W_{50})^2/(a_r)^2+C_6/a_r,\nonumber
\end{eqnarray}

\noindent where $W$ is a corrected HI line width in km/s measured
at 50{\%} of maximum, $a_{r}$ and $a_{b}$ are corrected major
galaxies' angular diameters on POSS and ESO/SERC reproductions,
$T$ is a morphological type indicator ($T=I_{t}-5.35$, where
$I_{t}$ is a Hubble type; $I_{t}=5$ corresponds to type Sc) and
$B$ is a surface brightness indicator ($B=I_{SB}-2$, where
$I_{SB}$ is a surface brightness index from RFGC; brightness
decreases from I to IV).

The D-model has 9 parameters (6 coefficients $C$ and 3 components
of vector $\vec{D}$), DQ-model has 14 parameters (5 components of
tensor $\bf{Q}$ are added) and DQO-model are described by 24
coefficients. They was calculated by least square method for
different subsamples with distances limitation to make the sample
more homogeneous in depth (Parnovsky and Tugay, 2004). For preparing
this list we used coefficients for subsample with $R\le 1000\,km/s$ given
out in (Parnovsky and Tugay, 2004).

Note that there are other models of collective motion based on more sophisticated general
relativistic approach. In the paper by Parnovsky and Gaydamaka (2004) they were applied to
the sample mentioned above. Using coefficient obtained in this paper and data from our Table 1
one can make a list of peculiar velocities for relativistic and semirelativistic models of
galaxies' motion.

\section{Samples}
Observational data were divided into several samples.
\begin{enumerate}
\item The observations of flat galaxies from FGC were performed with the 305\,m telescope at Arecibo
(Giovanelli et al., 1997). The observations are confined within
the zone $0\,^{\circ}<\delta\le+38\,^{\circ}$ accessible to the radio
telescope. There was no selection by the visible angular diameter,
type, axes ratio and other characteristics. We have not included
in the summary the flat galaxies from the Supplement to FGC, which
do not satisfy the condition $a/b\ge7$, and also the galaxies with
uncertain values of $W_{50}$, in accordance with the notes in the
paper by Giovanelli et al. (1997). Our list contains 486 flat
galaxies from this paper.
\item The observations of optical rotational curves made with the 6\,m
telescope of SAO RAS (Makarov et al., 1997\,a,\,b; 1999; 2001).
The objects located in the zone $\delta \ge 38\,^{\circ}$, with the axes
ratio $a/b \ge 8$ and a large diameter $a\le 2'$ were selected for
the observations. The maximum rotational velocities were converted
to $W_{50}$ by a relation derived through comparison of optical
and radio observations of 59 galaxies common with sample ``1''
(Makarov et al., 1997a). 286 galaxies from these papers are
included into our list.
\item The data on radial velocities and hydrogen line widths in the FGC
galaxies identified with the RC3 catalogue (de Vaucouleurs et al., 1991).
In a few cases, where only $W_{20}$ are available in RC3, they were converted
to $W_{50}$ according to Karachentsev et al. (1993). This sample comprises
flat galaxies all over the sky, a total of 162 objects.
\item The data on HI line widths (64\,m radio telescope, Parkes) and
on optical rotational curves $V_{rot}$ (2.3\,m telescope of Siding
Spring) for the flat galaxies identified with the lists by
Mathewson et al. (1992), Mathewson and Ford (1996). The optical
data were converted to the widths $W_{50}$ according to Mathewson
and Ford (1996). The Sb--Sd galaxies from the catalogue
ESO/Uppsala (Lauberts, 1982) with angular dimensions $a\ge 1'$,
inclinations $i>40\,^{\circ}$, and a galactic latitude $(|b|)\ge 11\,^{\circ}$
have been included in the lists.  As Mathewson et al. (1992)
report, the data obtained with the 64\,m and 305\,m telescopes are
in good agreement. Our sample contains 166 flat galaxies from
these papers.
\item The HI line observations of flat galaxies carried out by Matthews and van Driel (2000) using
the radio telescopes in Nancay ($\delta > -38\,^{\circ}$) and Green Bank
($\delta =-38\,^{\circ} \div -44.5\,^{\circ}$). They have selected the flat
galaxies from FGC(E) with angular dimensions $a>1'$, of Scd types
and later, mainly of low surface brightness (SB\,=\,III and IV
according to RFGC). We did not include in our list uncertain
measurements from the data of Matthews and van Driel (2000). In
the case of common objects with samples ``1'' or ``2'' we excluded
the data of Matthews and van Driel (2000) on the basis of the
comparisons of $V_h$ and $W_{50}$. The subsample ``5'' comprises
194 galaxies.
\item Data from the HyperLeda extragalactic database. This sample
includes 233 new entries in comparison with (Karachentsev et al, 2000) and new
data for 34 galaxies listed in (Karachentsev et al, 2000).
\end{enumerate}

\section{List of peculiar velocities description}

These models were applied to the computation of peculiar
velocities of all 1561 galaxies. They are presented in Tables\,1
and 2. The content of the columns in Table 1 is as follows: \\
(1), (2) --- the number of the galaxy in the RFGC and FGC
catalogues, respectively; \\ (3) --- the right ascension and
declination for the epoch 2000.0; \\ (4), (5) --- the corrected
``blue'' and ``red'' major diameters, in arcmin; \\ (6) --- the
corrected line width $W_{50}$ in km/s; \\ (7) --- the radial
velocity in the system of 3K cosmic microwave radiation, in km/s;
\\ (8) --- the number of the sample from which the original data
$V_h$ and $W_{50}$ were taken. A ``B'' note after this number
means that this is a ``bad'' data. A ``N'' note means that this
galaxy is a one from 34 galaxies which data were changed in
comparison with the previous list (Karachentsev et al, 2000).\\
Columns (9) -- (13)
contain a peculiar velocities list for D-model:\\ (9) --- the
distance (in km/s) measured from the basic regression on the
assumption that the model of motion of galaxies is dipole; \\ (10)
--- the dipole component of the radial velocity, in km/s; \\
(11) --- the value of radial velocity (in km/s) from regression
(1): $V^{reg}=Hr+V^{dip}$;\\ (12) --- the deviation of radial
velocity from regression (1), in km/s: $\delta
V=V_{3K}-V^{reg}$;\\ (13) --- the peculiar velocity, in km/s:
$V^{pec}=V_{3K}-Hr$.

The details about correction one can see in (Karachentsev et al, 2000).

In the Table 2 we present data for DQ- and DQO-models. The content
of the columns in it is as follows: \\ (1) --- the number of the
galaxy in the RFGC catalogue; \\ (2) --- the distance (in km/s)
for DQ-model;\\ (3), (4) --- the dipole and quadrupole radial
components of galaxies' large-scale motion for DQ-model;\\ (5) ---
the radial velocity for DQ-model (in km/s) from regression (2):
$V^{reg}=Hr+V^{dip}+V^{qua}$;\\ (6) --- the deviation of radial
velocity from regression (2), in km/s: $\delta
V=V_{3K}-V^{reg}$;\\ (7) --- the peculiar velocity for DQ-model,
in km/s: $V^{pec}=V_{3K}-Hr$; \\ (8) --- the distance (in km/s)
for DQO-model;\\ (9), (10), (11) --- the dipole, quadrupole and
octopole radial components of galaxies' large-scale motion for
DQO-model;\\ (12) --- the radial velocity for DQO-model (in km/s)
from regression (3): $V^{reg}=Hr+V^{dip}+V^{qua}+V^{oct}$;\\ (13)
--- the deviation of radial velocity from regression (3), in km/s:
$\delta V=V_{3K}-V^{reg}$;\\ (14) --- the peculiar velocity for
DQO-model, in km/s: $V^{pec} =V_{3K}-Hr$. \\

An ASCII file with the data from Tables 1 and 2 with some additional columns containing
indices of type and surface brightness class can be obtained (naturally, free of charge)
by e-mail request to par@observ.univ.kiev.ua with subject ``list''.

Note that the data from this list were already used to obtain a
density distribution up to $80h^{-1}$ Mpc and estimation of
cosmological parameters $\Omega_m$ and $\sigma_8$, corresponding
papers are submitted.

{}
%
%\vspace{1cm}

%\clearpage
%\newpage
%\newpage

\begin{center}
%\mbox{\hspace*{15cm}}\\
\topcaption{A list of velocity--distance data for the RFGC galaxies}
\tablehead{
\hline
\multicolumn{1}{c}{RFGC}&
\multicolumn{1}{c}{FGC}&
\multicolumn{1}{c}{RA (2000) D}&
\multicolumn{1}{c}{$a_b$}&
\multicolumn{1}{c}{$a_r$}&
\multicolumn{1}{c}{$W_{50}$}&
\multicolumn{1}{c}{$V_{3K}$}&
\multicolumn{1}{c}{S}&
\multicolumn{1}{c}{Hr}&
\multicolumn{1}{c}{$V^{dip}$}&
\multicolumn{1}{c}{$V^{reg}$}&
\multicolumn{1}{c}{$\delta V$}&
\multicolumn{1}{c}{$V^{pec}$} \\ \hline}
\tabletail{\hline}
\par
\begin{supertabular}{rrrrrrrrrrrrr}\hline
   1&2565  &000056.1$+$202017&1.33&1.27&416.3& 6452&1  & 6510&$-$233& 6278&  174& $-$58\\
   4&2567  &000136.8$+$033017& .94& .88&251.5& 5982&1  & 6059&$-$180& 5879&  104& $-$76\\
   6&2563  &000204.3$+$803832& .98& .92&228.8& 4002&2  & 4682&$-$244& 4438&$-$437&$-$681\\
  12&2573  &000223.3$+$271238&1.19&1.00&292.3& 7291&2  & 5461&$-$249& 5211& 2079& 1830\\
  14&   1  &000403.0$-$065845& .56& .53& 94.9& 3375&5  & 2483&$-$140& 2343& 1032&  892\\
  22&   6\E&000541.0$-$485848& .51& .45&120.7& 6411&5  & 4392&  55& 4447& 1965& 2020\\
  26&   5  &000638.1$+$171703& .82& .86& 78.4&  529&6  & 1273&$-$226& 1047&$-$517&$-$744\\
  27&   6  &000653.7$+$414426&1.09& .92&254.3& 3951&2  & 5103&$-$272& 4831&$-$880&$-$1152\\
\end{supertabular}
\end{center}

\begin{center}
\topcaption{A list of velocity--distance data for the RFGC galaxies according to DQ- and DQO-models}
\tablehead{%
\hline
\multicolumn{1}{c}{No}&
\multicolumn{1}{c}{$Hr_q$}&
\multicolumn{1}{c}{$V^{dip}$}&
\multicolumn{1}{c}{$V^{qua}$}&
\multicolumn{1}{c}{$V^{reg}$}&
\multicolumn{1}{c}{$\delta V$}&
\multicolumn{1}{c}{$V^{pec}$}&
\multicolumn{1}{c}{$Hr_o$}&
\multicolumn{1}{c}{$V^{dip}$}&
\multicolumn{1}{c}{$V^{qua}$}&
\multicolumn{1}{c}{$V^{oct}$}&
\multicolumn{1}{c}{$V^{reg}$}&
\multicolumn{1}{c}{$\delta V$}&
\multicolumn{1}{c}{$V^{pec}$}\\ \hline}
\tabletail{\hline}
\par
\begin{supertabular}{rrrrrrrrrrrrrr}\hline

   1& 6497&$-$190& $-$90& 6217&   235&  $-$45& 6515&$-$147& $-$47&  143& 6464&  $-$12&  $-$63\\
   4& 6081&$-$128&   67& 6021&  $-$38&  $-$99& 6078& $-$98&  128&  244& 6352& $-$369&  $-$96\\
   6& 4669&$-$259& $-$90& 4320& $-$318& $-$667& 4662&$-$202& $-$82& $-$26& 4351& $-$350& $-$660\\
  12& 5472&$-$212&$-$121& 5138&  2152&  1819& 5469&$-$164& $-$93&   60& 5271&  2020&  1822\\
  14& 2436& $-$83&   64& 2418&   958&   939& 2417& $-$64&   92&   46& 2491&   884&   958\\
  22& 4374&  109&  196& 4679&  1733&  2038& 4335&   86&  242&   79& 4741&  1671&  2077\\
  26& 1242&$-$181& $-$10& 1050& $-$521& $-$712& 1222&$-$142&  $-$1&    6& 1085& $-$556& $-$693\\
  27& 5111&$-$247&$-$176& 4688& $-$737&$-$1160& 5103&$-$194&$-$164&  $-$8& 4737& $-$786&$-$1152\\
\end{supertabular}
\end{center}
\end{document}